\documentclass[12pt]{article}

\usepackage{epsf}
\usepackage{amssymb}
\usepackage{cite}

\def\ee{\varepsilon}
\def\mprp{\mbox{\tiny $\bot$}}
\def\mprl{\mbox{\tiny $\|$}}

\title{
\begin{flushright}
{\normalsize Yaroslavl State University\\
             Preprint YARU-HE-02/06\\
             hep-ph/0209253} \\[10mm]
\end{flushright}
Could the process \\
of neutrino photoproduction on nuclei, \\
stimulated by a strong magnetic field, \\
compete with URCA processes?}
 
\author{
A. V. Kuznetsov and N. V. Mikheev\\[3mm]
{\small\it Division of Theoretical Physics,} \\
{\small\it Yaroslavl State (P.G.~Demidov) University,} \\
{\small\it Sovietskaya 14, 150000 Yaroslavl, Russian Federation}\\
{\small\tt E-mail: avkuzn@uniyar.ac.ru, mikheev@uniyar.ac.ru}
}

\date{}

\begin{document}

\maketitle

\begin{abstract}
The recent studies~\cite{Kuznetsov:2002} is reported of the 
neutrino photoproduction on nuclei,
$\gamma + Ze \to Ze + \gamma + \nu + \bar\nu$,
in a strong magnetic field.
It is shown that the catalyzing influence of the field on the 
process decreases essentially because of the modification 
of the photon dispersion properties in a strong magnetic field.
Therefore, at any field magnitude, neutrino photoproduction 
cannot compete with the URCA processes. 
This conclusion contradicts a recent statement in the 
literature~\cite{Skobelev:2001}.
\end{abstract}
 
\vfill

\begin{center}
{\sl Talk presented at the 12th International Seminar \\
``Quarks-2002'', \\
Valday and Novgorod, Russia, June 1-7, 2002}
\end{center}

\newpage

\section{Introduction}

Strong magnetic fields which could be generated in the astrophysical 
cataclysms like a supernova explosion or a coalescence of neutron stars, 
make an active influence 
on quantum processes, thus allowing or enhancing the transitions 
which are forbidden or strongly suppressed in vacuum.
However, the magnetic field influences significantly the quantum processes 
only in the case when it is strong enough. 
There exists a natural scale for the field strength which is the so-called 
critical value $B_e = m_e^2/e \simeq 4.41 \cdot 10^{13}\,$ G
(we use natural units in which $c = \hbar = 1$, hereafter 
$e$ is the elementary charge). 

There exist arguments that field of such and essentially greater scale
can appear in astrophysical objects. Thus, a class of stars exists, 
the so-called magnetars, which are neutron stars with magnetic fields 
$\sim 4 \cdot 10^{14}\,$ G~\cite{Kouveliotou:1999,Hurley:1999}.
Models of astrophysical processes and objects are discussed, 
where magnetic fields of the order $10^{17} - 10^{18}\,$ G can be 
generated~\cite{Bisnovatyi-Kogan:1970,Duncan:1992,Bocquet:1995,Cardall:2001}.
Thereby, physics of quantum processes in strong external fields presents 
itself as an interesting and important direction of studies, both from 
a conceptual standpoint, and in light of possible astrophysical 
implications.

Among others, the set of quantum processes is very interesting where only 
electrically neutral particles, such as neutrinos and photons, are presented 
in the initial and final states. The external field influence on these 
loop processes is provided, first, by the sensitivity of the charged 
virtual fermion to the field, and the electron plays the main role here 
as the particle with the maximal specific charge, $e / m_e$. 
Secondly, strong magnetic field essentially influences the dispersion 
properties of photons, and consequently it changes their kinematics.

In the recent paper~\cite{Skobelev:2001} a contribution was studied, 
in particular, of the loop process of the neutrino pair photoproduction 
on nucleus
\begin{eqnarray}
\gamma + Ze \to Ze + \gamma + \nu + \bar\nu
\label{eq:reac1}
\end{eqnarray}
in strong external magnetic field, into the star cooling. 
\begin{figure}[ht]
\hspace{30mm}
\epsfxsize=0.5\textwidth
\epsfbox{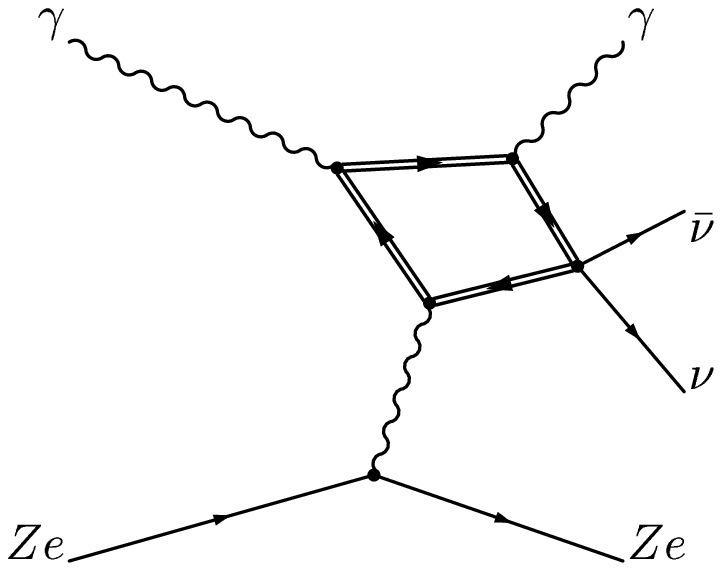} 
\caption{\small 
The Feynman diagram for the neutrino pair photoproduction 
on nucleus in a magnetic field.}
\label{fig:photopr}
\end{figure}
An important conclusion was made there, that a contribution of this 
process could compete with the contribution of URCA - processes. Thereby, 
the process~(\ref{eq:reac1}) should be taken into account 
in the description of a cooling of strongly magnetized neutron star, 
as one more channel of neutrino energy losses.

Here, we present the result of a new study of the process of neutrino pair 
photoproduction on nucleus, Fig.~\ref{fig:photopr}. 
We show that with the dispersion 
of a photon in strong magnetic field taken into account, 
a catalizing influence of the field on the process~(\ref{eq:reac1}) 
vastly decreases. This effect was not considered in 
Ref.~\cite{Skobelev:2001} with the result that the contribution of 
the loop process turned out to be overestimated in many orders of 
magnitude.

\section{The process $\gamma + \gamma + \gamma \to \nu + \bar\nu$\\
in a strong magnetic field}

We start from this photon - neutrino process
which is symmetric with respect to the photon interchange.
The crossed process having a physical meaning is the 
photon - neutrino process $\gamma \gamma \to \nu \bar\nu \gamma$, 
and the history of its investigations is rather long.
It was studied in vacuum by Van Hieu and Shabalin~\cite{VanHieu:1963} 
and by Dicus and Repko~\cite{Dicus:1997}. 
The process was studied in a strong magnetic field ($B \gtrsim B_e$)
in the paper~\cite{Loskutov:1987}.

The amplitude of the process in a strong field has the form
\begin{eqnarray}
{\cal M} &=& - \frac{8\,e^3\,G_F\,e B}{\sqrt{2} \pi^2\,m_e^6}\,
(\ee_1 \tilde\varphi k_1)\;
(\ee_2 \tilde\varphi k_2)\;
(\ee_3 \tilde\varphi k_3)\;
\times
\nonumber\\
&\times&
[C_V\,(j \tilde\varphi k_4) + C_A (j \tilde\varphi \tilde\varphi k_4)]
I \left(\frac{k_1}{m_e},\frac{k_2}{m_e},\frac{k_3}{m_e}\right),
\label{eq:ampl}
\end{eqnarray}
where
$C_V,\;C_A$ are the vector and axial-vector constants of the effective
$\nu \nu e e$ Lagrangian,
\begin{eqnarray}
C_V = \pm 1/2 + 2 \sin^2 \theta_W, \; C_A = \pm 1/2, 
\label{eq:CVCA}
\end{eqnarray}
(here the upper signs correspond to $\nu_e$, and the lower signs 
correspond to $\nu_\mu$ and $\nu_\tau$);
$\ee_{1,2,3}$ and $k_{1,2,3}$ are the polarization 4-vectors and the 
momenta of photons,
$j_\alpha = [\bar\nu(q_1) \gamma_\alpha (1 + \gamma_5) \nu(-q_2)]$ 
is the Fourier transform of the neutrino current,
$k_4 = q_1 + q_2$ is the neutrino pair momentum,
$\tilde\varphi_{\alpha \beta} = {\tilde F}_{\alpha \beta} /B$ 
is the dimensionless dual tensor of external magnetic field,
${\tilde F}_{\alpha \beta} = \frac{1}{2} \varepsilon_{\alpha \beta
\mu \nu} F_{\mu \nu}$. 
The tensor indices of four-vectors and tensors standing inside the 
parentheses are contracted consecutively, for example
$(a \tilde \varphi b) = a_\alpha {\tilde \varphi}_{\alpha \beta} b_\beta$.

The formfactor $I \left(\frac{k_1}{m_e},\frac{k_2}{m_e},\frac{k_3}{m_e}
\right)$ has a complicated form of a triple 
integral over the Feynman variables. 
In the case of low photon energies, $\omega_{1,2,3} \ll m_e$, 
the integral can be easily calculated: 
\begin{eqnarray}
I \left(\frac{k_1}{m_e},\frac{k_2}{m_e},\frac{k_3}{m_e}\right)
\simeq \frac{1}{60}.
\label{eq:I_k_low}
\end{eqnarray}
In this case, the above amplitude corresponds to the effective local 
Lagrangian of the $\gamma \gamma \gamma \nu \bar\nu$ interaction:
\begin{eqnarray}
{\cal L}_{eff} &=& 
- \frac{e^3\,G_F\,e B}{45\,\sqrt{2} \pi^2\,m_e^6}\,
\left(\frac{\partial A^\alpha}{\partial x_\beta}\;
\tilde\varphi_{\alpha \beta} \right)^3
\times
\nonumber\\
&\times&
\frac{\partial}{\partial x_\sigma} 
[\bar\nu \gamma^\rho (1 + \gamma_5) \nu ]\;
[C_V\, \tilde\varphi_{\rho \sigma} + C_A \,
(\tilde\varphi \tilde\varphi)_{\rho \sigma}].
\label{eq:L_eff}
\end{eqnarray}
We note that $\gamma \gamma \gamma \nu \bar\nu$ interaction in the 
low-energy limit was studied earlier by Loskutov and 
Skobelev~\cite{Loskutov:1987}, 
however, their Lagrangian has an extra factor of 2.

The dimensional analysis of the amplitude with respect to a typical 
photon energy, $|k_1| \sim |k_2| \sim |k_3| \sim \omega$, shows 
an essential distinction of the cases of low energies 
(${\cal M} \sim \omega^5$), and high energies, (${\cal M} \sim \omega^{-3}$). 
 
\section{Photon dispersion and kinematics \\
in a strong magnetic field}

In analyses of the photon processes in a strong magnetic field, 
the field influence on the photon dispersion properties is a crucial 
factor, and it has to be taken into account. 
We remind that only photons of transversal polarization~\cite{Adler:1971} 
participate in the processes in a strong magnetic field.
For virtual photon, instead of the propagator $\sim q^{-2}$, 
one should use 
the propagator with the vacuum polarization in a magnetic field:
\begin{eqnarray}
D^{(B)} (q_{\mprl}^2,q_{\mprp}^2) \; = \; \frac{1}{q^2 - P(q_{\mprl}^2)},
\label{eq:prop}
\end{eqnarray}
here $q_{\mprl}^2 = q_0^2 - q_z^2, \; q_{\mprp}^2 = q_x^2 + q_y^2, \; 
q^2 = q_{\mprl}^2 - q_{\mprp}^2$ (the magnetic field is directed 
along the $z$ axis),
$P(q_{\mprl}^2)$ is the photon polarization operator in the field, 
which has a rather simple form in a strong field, 
$B \gg B_e$, and in an approximation $|q_{\mprl}^2| \ll m_e^2$ 
\cite{Shabad:1988}:
\begin{eqnarray}
P(q_{\mprl}^2) \simeq - \frac{\alpha}{3 \pi} \; \frac{B}{B_e} \;q_{\mprl}^2.
\label{eq:polar1}
\end{eqnarray}
It is convenient to introduce a dimensionless parameter which 
defines the field influence in expressions below:
\begin{eqnarray}
\beta = \frac{\alpha}{3 \pi} \; \frac{B}{B_e}.
\label{eq:beta}
\end{eqnarray}
For the field values $10^3\, B_e$ and $10^4 \, B_e$, 
the parameter $\beta$ is 0.77 and 7.7 consequently, so, 
it is not the small one.

Finally, taking $q_0 = 0$ for a virtual photon coupled with a nucleus at rest, 
one obtains the propagator:
\begin{eqnarray}
D^{(B)} \; \simeq \; - \; \frac{1}{q_{\mprp}^2 + (1 + \beta) q_z^2}.
\label{eq:prop1}
\end{eqnarray}

On the other hand, real photons participating in the process are also 
under the influence of a strong magnetic field, which leads to 
the renormalization of the photon wave-functions:
\begin{eqnarray}
\ee_\alpha \longrightarrow \sqrt{\cal Z}\,\ee_\alpha, 
\label{eq:eer}
\end{eqnarray}
where the renormalization factor $\cal Z$ 
takes the form
\begin{eqnarray}
{\cal Z} = \left( 1 - \frac{d P (q_{\mprl}^2)}{d q_{\mprl}^2} \right)^{-1} 
= \frac{1}{1 + \beta}.
\label{eq:z}
\end{eqnarray}
The kinematics of photons is also modified by the field.
The photon dispersion equation $k^2 - P(k_{\mprl}^2) = 0$ 
can be rewritten to the form
$\omega^2 = {\bf k}^2 (1 + \beta \cos^2 \theta)/(1 + \beta)$, 
and the element of the momentum space becomes
$$d^3 k = (1 + \beta)\, \omega^2 d \omega \, d y \, d \varphi, 
\quad y = \cos \theta \sqrt{1 + \beta} \big / \sqrt{1 + \beta \cos^2 
\theta},$$
where $\theta, \varphi$ are the polar and the azimuthal angles.

\section{The process of neutrino photoproduction \\ on nuclei}

Using the effective local Lagrangian of the 
$\gamma \gamma \gamma \nu \bar\nu$ interaction, 
with the field influence on the photon properties, 
and with the substitution of the photon $\bot$ polarization vectors
\begin{eqnarray}
\ee_\alpha^{(\mprp)} = 
\sqrt{\cal Z}\,\frac{(\tilde \varphi k)_\alpha}{\sqrt{k_{\mprl}^2}},
\label{eq:eps}
\end{eqnarray}
the amplitude for the process
$$\gamma + Ze \to Ze + \gamma + \nu + \bar\nu$$
can be presented in the form
\begin{eqnarray}
{\cal M} = \frac{32 \pi \alpha Z\,G_F}{5\,\sqrt{2}\,m_e^4}\;
\frac{\beta}{1 + \beta} \;
\frac{2 m_N\,q_z \sqrt{k_{1 \mprl}^2 k_{2 \mprl}^2}}
{q_{\mprp}^2 + (1 + \beta) q_z^2} \;
[C_V\,(j \tilde\varphi k_4) + C_A (j \tilde\varphi \tilde\varphi k_4)],
\label{eq:ampl2}
\end{eqnarray}
where $m_N$ is the nucleus mass, $q$ is the momentum transferred,
$q^\alpha = (0, {\bf q})$. 

Our amplitude differs essentially from the amplitude obtained 
in the paper~\cite{Skobelev:2001},
where the strong magnetic field influence 
on the photon dispersion properties was not taken into account.

\section{The neutrino emissivity}

The neutrino emissivity is the energy carried out by neutrinos 
from unit volume per unit time. It is defined in terms of the process 
amplitude~(\ref{eq:ampl2}) as follows
\begin{eqnarray}
Q_\nu &=& \frac{(2 \pi)^4 n_N}{2 m_N} \int |{\cal M}|^2 \;
(\ee_1 + \ee_2) \;
\delta^4 (k_1 - k_2 - q_1 - q_2 - q) 
\times
\nonumber\\
&\times&
\frac{d^3 k_1}{(2 \pi)^3 2 \omega_1} f (\omega_1)
\frac{d^3 k_2}{(2 \pi)^3 2 \omega_2} [1 + f (\omega_2)]\,
\times
\label{eq:Q1}\\
&\times&
\frac{d^3 q_1}{(2 \pi)^3 2 \ee_1}\, \frac{d^3 q_2}{(2 \pi)^3 2 \ee_2}\, 
\frac{d^3 q}{(2 \pi)^3 2 m_N}, 
\nonumber
\end{eqnarray}
where $n_N$ is the nuclei density,
$\ee_1$ and $\ee_2$ are the neutrino and antineutrino energies,
$f (\omega) = [\exp(\omega/T) - 1]^{-1}$
is the density of the photon gas in equilibrium at the temperature $T$. 
One obtains
\begin{eqnarray}
Q_\nu = \frac{8\,(2 \pi)^9}{225}\;Z^2\,\alpha^2\,G_F^2 \,m_e^6\,n_N\,
\left(\frac{T}{m_e}\right)^{14}\;J\,(\beta).
\label{eq:Q2}
\end{eqnarray}
The dependence of the value $Q_\nu$ on the field parameter 
$\beta$~(\ref{eq:beta}) 
is defined by the function $J\,(\beta)$ as follows
\begin{eqnarray}
J\,(\beta) &=& \beta^2\, \int\limits_{-1}^1 du \,(1 - u^2)
\int\limits_{-1}^1 dv \,(1 - v^2) \int\limits_0^1 ds \,s^3 \,(1 - s)^8 
\int\limits_0^1 dr \,r^2
\times
\nonumber\\
&\times&
\int\limits_{-1}^1 dx [u - s v - (1 - s) r x]^2
(1 - r^2 x^2) 
\times
\label{eq:J}\\
&\times&
\left[\overline{C_V^2} (1 - r^2) 
+ \overline{C_A^2} r^2 (1 - x^2)\right]
\int\limits_0^{2 \pi} \frac{d \varphi_1}{2 \pi}
\int\limits_0^{2 \pi} \frac{d \varphi_2}{2 \pi} \;\frac{1}{[F (\beta)]^2},
\nonumber
\end{eqnarray}
\begin{eqnarray}
F (\beta) &=& (1 + \beta) \left\{1-u^2 +s^2 (1-v^2) 
- 2 s \sqrt{1-u^2} \sqrt{1-v^2} \cos \varphi_1 
+ \right .
\nonumber\\
&+& 
\left . [u - s v - (1 - s) r x]^2 \right\} 
- 2 \sqrt{1 + \beta} (1-s) r \sqrt{1-x^2} 
\times
\nonumber\\
&\times& 
\left[\sqrt{1-u^2}\cos \varphi_2
- s \sqrt{1-v^2}\cos (\varphi_2 - \varphi_1)\right] 
+
\nonumber\\
&+&
(1-s^2) r^2 (1-x^2).
\label{eq:F}
\end{eqnarray}
The constants $\overline{C_V^2} = 0.93$ and $\overline{C_A^2} = 0.75$ 
under the integral are summarized over all channels of the 
neutrino production, $\nu_e, \nu_\mu, \nu_\tau$. 

\begin{figure}[ht]
\hspace{10mm}
\epsfxsize=0.8\textwidth
\epsfbox{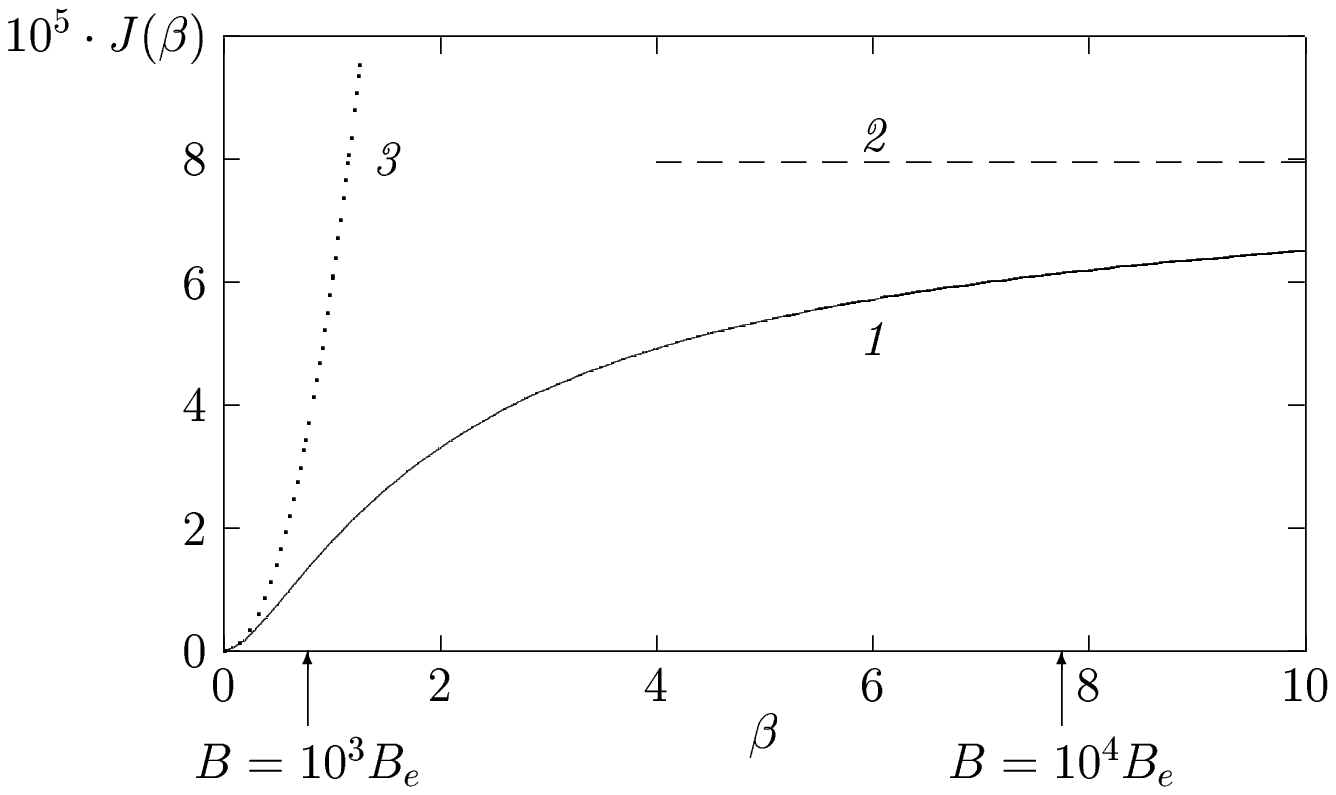} 
\caption{\small 
The dependence of the function $J (\beta)$ on the field 
parameter $\beta$ (the curve {\it 1}). The line {\it 2} shows
the asymptotics of the function at large values of $\beta$, 
$J (\beta) \to 8 \cdot 10^{-5}$. 
The curve {\it 3} shows the dependence $\sim \beta^2$, 
which would take place without taking account of the magnetic field 
influence on the photon dispersion.}
\label{fig:j_beta}
\end{figure}

The dependence of the function $J (\beta)$ on the field 
parameter $\beta$ is shown in Fig.~\ref{fig:j_beta}. 

The upper bound for the value $Q_\nu$ in the asymptotically 
strong field is
\begin{eqnarray}
Q_\nu \lesssim 2.3 \cdot 10^{27} \;
\left(\frac{T}{m_e}\right)^{14}
\left <\frac{Z^2}{A}\right> \; \left(\frac{\rho}{\rho_0}\right) \;
\frac{\mbox{erg}}{\mbox{cm}^3 \;\mbox{s}},
\label{eq:Q3}
\end{eqnarray}
where $Z$ is the charge number and $A$ is the mass number of a nucleus, 
\newline
$<\!{Z^2}/{A}\!>$ means the averaging over all nuclei,
$\rho_0 = 2.8 \cdot 10^{14} \, \mbox{g}/\mbox{cm}^3$
is the typical nuclear density, and $\rho$ is the averaged density 
of a star.

The Eq.~(\ref{eq:Q3}) should be compared with the neutrino 
emissivity via the standard channel of the modified URCA process
\begin{eqnarray}
Q_\nu(\mbox{URCA}) \sim 10^{27} \;
\left(\frac{T}{m_e}\right)^{8}
\; \left(\frac{\rho}{\rho_0}\right)^{2/3} \;
\frac{\mbox{erg}}{\mbox{cm}^3 \;\mbox{s}}.
\label{eq:Q4}
\end{eqnarray}

At first glance, these values for the emissivities could 
compete. However, a big numerical factor in the 
neutrino photoproduction emissivity arises 
from the integral over the initial photon energy
$\omega_1$ ($x = \omega_1/T$)
\begin{eqnarray}
\int\limits_0^\infty \frac{x^{13}\;d x}{e^x - 1} \;=\; 13! \; \zeta (14)
\;=\; \frac{(2 \pi)^{14}}{24} \;\simeq\; 6.2 \cdot 10^9.
\label{eq:int}
\end{eqnarray}
It is obvious, that the integral~(\ref{eq:int}) acquires its big value 
in the region of the argument 
$$x \sim 10 \div 20, \quad \omega_1 \sim (10 \div 20) \,T.$$ 
Thus, as the amplitude of the neutrino photoproduction is obtained 
within the approximation 
$$\omega \lesssim m_e,$$
the above expression 
for the neutrino emissivity is true at the photon gas temperatures
$$T \lesssim (1/10) \,m_e,$$ 
however, it is obviously incorrect at $T \sim m_e$.

Consequently, the assumption that the factor $(T / m_e)^{14}$ 
could be taken of the order unity~\cite{Skobelev:2001}, is wrong. 
Within the area of applicability one obtains
$$(T / m_e)^{14} \lesssim 10^{-14}.$$

\medskip

In summary, the neutrino photoproduction on nuclei 
cannot compete with URCA processes for any values 
of the magnetic field strength.

\bigskip

{\bf Acknowledgements}  

We express our deep gratitude to the organizers of the 
Seminar ``Quarks-2002'' for warm hospitality.

This work was supported in part by the Russian Foundation for Basic 
Research under the Grant No.~01-02-17334
and by the Ministry of Education of Russian Federation under the 
Grant No.~E00-11.0-5.

\end{document}